# Pairwise Interactions Origin of Entropy Functions

**Yuri A. Pykh**

Research Center for Interdisciplinary Environmental Cooperation of Russian Academy of Sciences (INENCO RAS), Saint-Petersburg, 191187, nab. Kutuzova, 14, Russia;
Email: inenco@mail.neva.ru, malkinapykh@yandex.ru;
Tel: +7 – 921 – 9034092

**Abstract:** In this paper we combine the three universalisms: pairwise interactions concept, dynamical systems theory and relative entropy analysis to develop a theory of entropy issues. We introduce two hypotheses concerning the structure and types properties of the system's entities interactions and derive generalized replicator dynamic equations. Then we construct energy-like and entropy-like Lyapunov-Meyer functions (LMF) for these equations. We show that energy-like LMF contains no information about the equilibrium of the system and is a substantial generalization of the Fisher's fundamental theorem of natural selection. If there are exists nontrivial equilibrium point for generalized replicator system then we construct entropy-like LMF for this system and prove that it is a relative entropy function or the function of information divergence. We prove that negative relative entropy is a convex function for a probability space and receive new distance measure between two probability distributions. Also we use Legendre-Donkin-Fenchel transformation for dual coordinates. As result we establish the set of next important links: nonlinear pairwise interaction → generalized Fisher (replicator) equations → Lyapunov-Meyer functions → relative entropy → distance measure → LDF-transformation → duality.

In particular it follows from these links that nonlinear pairwise interaction is the origin of all known entropy functions.

**Keywords**: pairwise interactions; Lyapunov functions; stability postulate; entropy functions; escort distribution

# 1. Introduction

*1.1. Entropy and Lyapunov functions*

The primary goal of this paper is to study the implications of Lyapunov functions theory for the analysis of nonlinear pairwise interactions replicator systems. We build two types of Lyapunov-Meyer functions, either as an energy-like or as an entropy-like functions (distance measures). The existence of the Lyapunov-Meyer functions allows us to construct practically all known entropy and relative-entropy functions

Our investigations are based mainly on the stability postulate stated by Chetaev in 1936 [1]

**Stability postulate.** *Stability, which is a fundamentally general phenomenon, apparently, must manifest itself in basic laws of nature in some way. If knowledge is constructed from the requirement of small deviations from nature, then scientific thinking must (or can) rely on some Lyapunov function V. Certainly, this function always exists according to the stability postulate.*

Many years later in 1967 stability postulate, in more narrow sense, was used by B.D.Coleman and V.J. Mizel in famous paper "Existence of Entropy as a Consequence of Asymptotic stability" [2].

In 1968, this postulate was stated mathematically by Meyer [3], who proved that, for dynamical systems whose limit sets consist of only isolated rest points or cycles, i.e., for Morse—Smale systems, there always exists a Lyapunov function, which increases (or decreases) on the set of wandering points of the system. Meyer following by S.Smale [4] suggested the term energy functions for such functions; however, it is more natural and convenient to refer to them as Lyapunov–Meyer functions. One of the best known examples of such function is entropy in Boltzmann's H-theorem.

Ilya Prigogine first pointed out [5] the importance of the relationship between Lyapunov functions and entropy: *"The positive time direction is associated with the increase of entropy. Let us emphasize the strong and very specific way in which the one-sidedness of time appears in the second law. According to its formulation it leads to the existence of a function having quite specific properties as expressed by the fact that for an isolated system it can only increase in time. Such function plays an important role in modern theory of stability as initiated by the classic work of Lyapunov. For this reason they are called Lyapunov functions (or functionals).*

*The entropy S is a Lyapunov function for isolated systems. As shown in all textbooks thermodynamic potentials such as the Helmholtz or Gibbs free energy are also Lyapunov functions for other "boundary conditions" (such as imposed values of temperature and volume).*

*In all these cases the system evolves to an equilibrium state characterized by the existence of a thermodynamic potential. This equilibrium state is an "attractor" for non-equilibrium states."*

For dynamical systems arising from physics the Lyapunov functions will typically have a thermodynamic interpretation (energy, entropy etc.) but its origin is not evident.

In an attempt to show how such an interpretation this ideas might proceed, we analyze below a class of macrosystems with nonlinear pairwise interactions for with an assumption of structure and type of interactions has consequences that there are exist an entropy-like and energy-like LMF functions.

*1.2. Pairwise Interactions in nature and science: a review and outlook:*

In Bialek and Ranganatan's recent paper: "Rediscovering the power of pairwise interactions" [6] is emphasized that the analysis of many objects the considered in physics, chemistry, biology, economy and other natural sciences can be carried out on the basis of research of the interactions. The mathematical background of this idea was developed in the works [7,8] where on the basis of the well-known theorem of Kolmogorov-Arnold of representation of any continuous function by superposition of functions of one or two variables, showed that pairwise interactions possess the fundamental properties defining dynamics of processes in many physical tasks. In particular V. Kreinovich [8] have shown that: "An arbitrary particle interaction can be represented in pairwise form by adding a finite number of extra particles." Supporting this general expectation we will carry out the small chronological review of the most known models of the pairwise interactions.

• **Newton's law of gravitation**. Most physical theories such as gravity or thermodynamics are based on pairwise interactions between particles. Thus, for example, the law of universal gravitation opened by I. Newton in **1684**. According this law a force acting on the i-th body is equal to the sum of the corresponding pairwise forces. It is necessary to emphasize that this approach known under the name gravitational, is widely used at creation of models of dynamics of the most various interacting objects now [9].

• **Boltzmann's kinetic equation**. Boltzmann's equation – the main equation in the kinetic theory of gases (**1872**). It describes change of distribution function of molecules on speeds and on coordinates. In approach of a hypothesis of pairwise collisions this equation is reduced to one closed equation for one-partial function of distribution – to the kinetic equation of Boltzmann. For this equation Boltzmann proves well-known H – theorem. Using the modern language it is possible to tell that he found for this equation a Lyapunov function related to entropy. We would note also that using Boltzmann's ideas in 1963 T. Hill [10], constructed thermodynamics of "small" systems which received further the name of nanothermodynamics [11].

• **Equation of Lotka-Volterra**. In **1920** A.Lotka [12] at the description of chemical reactions, kinetics and then in **1926** V. Volterra [13] for modeling dynamics of the interacting groups of biological community used a hypothesis of linear pairwise interactions that led to creation of well-known system of the ordinary differential equations with square non-linearities. For these equations two types of Lyapunov functions were found: of entropy-like type and energy-like functions. Recently, A. Wilson [14] established connection between the kinetic equations of Boltzmann and Lotka-Volterra system. In the monograph [15] author analyzes generalized Lotka-Volterra system with nonlinear pairwise interactions, the paper [16] provides a survey of recent activity in this issue.

• **Izing Model.** In **1924** E.Izing [17] offered the mathematical model of ferromagnetism in statistical mechanics based on a hypothesis of pairwise interactions between spins. Now this model of statistical mechanics to which research the huge number of publications is devoted found application in many areas of natural sciences [18].

• **Fischer's equations.** In **1930** R. Fischer [19], developing the mathematical theory of populations genetic structure evolution, offered the equations of dynamics of this structure. This system of the equations, also as well as Lotka-Volterra system, is founded on a hypothesis of linear pairwise interactions that is quite natural to the individuals breeding in population. For this system Fischer formulated based on enegy-like Lyapunov's function the evolutionary principle known as the main theorem of natural selection. Later, in 1972, Yu.A.Pykh [20] constructs for this system Lyapunov's entropy-like function. The perfect analysis of classical Fischer's equations with linear pairwise interactions is given in the monograph [15].

• **Equations of biological macromolecules evolution.** In **1971** M. Eigen [21] used the principle of pairwise interactions at creation the model of prebiological evolution explaining emergence of the ordered structures from the disorder. For a catalytic cycle with the constant general organization this approach gives the system of the equations close to Fischer's equations [15].

• **Evolutionary games**. In **1978** Taylor and Jonker [22] based on the concept of Evolutionarily Stable Strategy offered by Maynard Smith and using the principle of pairwise interactions constructed system of the equations the defining dynamics of the interacting populations depending on the strategy chosen by individuals. In this work for the first time there was using a name the replicator equations, entered in 1976 by Dawkins [23]. The good review on a current state of the theory of evolutionary games is given in [24,25,26].

• **Neural networks.** In **1982** Hopfield [27] offered model of the neural network based on pairwise interactions between neurons. He also found Lyapunov's function for this model. From the formal point of view this equations correspond to the generalized Lotka-Volterra model [16]. Since the work of J.J. Hopfield, neural networks have become one of the most important concepts in the sciences of interacting systems. For example one can see [28-31].

• **Biological networks.** In **2006** E.Schneidman with coauthors [32] showed that for the description of functioning retinal ganglion cells as they respond to natural movies has enough accounting of pairwise interactions. This work was based on use of Izing model and the principle of a maximum entropy. The papers by Roudi et al. [33,34] and Tkacik et al. [35] provide a survey of recent activity in biological networks.

• **Financial markets dynamics.** Recent results of Bury [36,37] have shown that the market microstructure capturing almost all of the available information in the data of stock markets does not involve higher order than pairwise interactions. Farmer in [38] discusses a set of differential equations that describe the flow of capital and pointed out that these equations are equivalent to the generalized Lotka-Volterra equations with pairwise interactions.

**Social interactions models.** Zhao et al. in [39] consider a general model of pairwise human social interaction consisting of N agents representing interacting in social network. Lou et al. in [40] analyze standard generalized additive models (GAMs) with pairwise interactions.

As we have briefly discussed above, different types of pairwise interactions have played an important role in the analysis of many natural phenomenon.

## 2. The Theoretical Results.

*2.1. Equations of macrosystems with nonlinear pairwise interactions.*

One of the important assumptions in classic Lotka-Volterra and replicator systems is that the functional responses are linear functional responses. Usually additive linear interactions models for predicting combined interactions effects cannot account for non-linearities in combined functional response introduced by non-trophic interactions Ayala et al [41]. Exponential and power law responses were used to model social dynamics by Beechman and Farnsworth [42]. The PhD Theses by Merrifield [43] provides in Chapter 5 a review on the concept of a force including nonlinear in different interacting systems.

Consider a macrosystem formed by a sufficiently large number $N$ of interacting objects. The classical definition of macrosystems is as follows: these are systems in which a chaotic behavior at the microlevel transforms into a deterministic behavior at the macrolevel. Suppose that, at a moment $t$, the macrosystem under consideration contains $n$ different types of objects and the number of objects of type $i$ is $x_i(t)$, where $i = 1,\ldots,n$, and $\sum_{i=1}^{n} x_i(t) = N(t)$. Consider the relative numbers $p_i(t) = x_i(t)/N(t)$ of various objects types. Obviously, $\sum_{i=1}^{n} p_i(t) = 1$, i.e., $\mathrm{p}(t) \in \sigma_p^n = \{\mathrm{p} \in \mathbb{R}^n : p_i \geq 0, i = 1,2,\ldots,n, \mathbf{e}^T \mathrm{p} = 1\}$, where $\sigma_p^n$ is the standard simplex in Euclidean $n$-space $\mathbb{R}^n$ and, $\mathbf{e}$ is the vector of ones. Thus, the state of such a macrosystem at each moment $t$ is determined by the vector $\mathrm{p}(t) = (p_i(t),\ldots,p_n(t))$.

We consider the systems in which only pairwise interactions occur during a short time interval $(t, t+\Delta t)$; in other words, those systems in which simultaneous interactions of more than two particles are impossible. Such a constraint is fairly natural and has found a recent confirmation in [6,32]. We make the following two assumptions about interactions between objects in a macrosystem [44].

**Hypothesis 1.** *The interaction between objects of types i and j is characterized by the so-called interaction strength, which we denote by $w_{ij}$ and regard as a quantitative characteristic of the effect of the interaction between two objects of types i and j on the rate of change of the relative number $p_i(t)$ of objects of type i.*

**Remark 1.** *Apparently, the notion of an interaction strength first appeared in mathematical ecology. A fairly detailed study of this notion and a survey of related results are contained in [45].*

**Remark 2.** *Note that the definition given above does not imply that $w_{ij} = w_{ji}$. We also emphasize that the asymmetry of results of interaction plays an essential role in many cases [46].*

**Hypothesis 2**. *For each macrosystem under consideration, there exists a set of probability distribution functions $f_i(p_i)$, $i = 1,...,n$, which determine the probability of the interaction of each object of type i with any other object in the macrosystems. Thus, the probability of the pairwise interaction between objects of types i and j is determined by the product $f_i(p_i) f_j(p_j)$.*

**Remark 3.** *The probability distribution functions $f_i$ (nonlinear response functions) obeys the conditions $f_i(0)=0$, $\partial f_i / \partial p_i > 0$ for $p_i > 0$, $\partial f_i / \partial p_i \geq 0$ for $p_i = 0$ and $f_i(1) = 1$;*

It follows from the two hypothesis stated above that, for small $\Delta t$, the dynamics of the system is determined by the relation

$$p_i(t+\Delta t) = p_i(t) + \Delta t \sum_{j=1}^{n} w_{ij} f_i(p_i(t)) f_j(p_j(t)) + o(\Delta t), \quad (1)$$

subject to the constraints $\sum_{i=1}^{n} p_i(t) - 1 = 0$, and $f_i(0) = 0$, for $i = 1, 2, ..., n$ which ensure the invariance of the simplex $\sigma_p^n$ and all of its faces. Using an approach common in analytic mechanics, we treat the system under consideration as a nonfree system subject to the ideal holonomic constraints

$$f_i(p_i)\left(\sum_{i=1}^{n} p_i(t) - 1\right) = 0, \quad i = 1,...,n \quad (2)$$

and apply the principle of replacing constraints by reaction forces to this system [47]. Passing to the limit as $\Delta t \to 0$ in (1) and taking into account the obvious identity $\sum \dot{p}_i = 0$ for the sum of derivatives, we obtain the following equation for the constraint multipliers (the indeterminate Lagrange multipliers) $\lambda_i$:

$$\sum_{ij}^{n} w_{ij} f_i(p_i) f_j(p_j) - \sum_{i=1}^{n} \lambda_i f_i(p_i) = 0$$

This equation has two obvious solutions, the trivial solution $\lambda_i = \sum_{j=1}^{n} w_{ij} f_j(p_j)$ and the nontrivial solution

$$\lambda_i = \lambda = \frac{\sum_{ij}^{n} w_{ij} f_i(p_i) f_j(p_j)}{\sum_{i=1}^{n} f_i(p_i)}, \quad i = 1,...,n$$

Using the nontrivial solution and denoting the sum $\sum_{i=1}^{n} f_i(p_i)$ by $\theta(\mathbf{p})$, we obtain the following system of differential equations determining the evolution of the probability distribution $\mathbf{p}(t)$:

$$\dot{p}_i = f_i(p_i)\left(\sum_{j=1}^{n} w_{ij} f_j(p_j) - \theta^{-1}(\mathbf{p})\sum_{ij}^{n} w_{ij} f_i(p_i) f_j(p_j)\right), \quad i=1,...,n \qquad (3)$$

This system of equations was first obtained from balance considerations in [48] without the use of the Lagrange method. In what follows, it is convenient to pass to the matrix form. System (3) is written in this form as

$$\dot{\mathbf{p}} = D(\mathbf{f})\left(\mathbf{Wf} - \mathbf{e}\theta^{-1}(\mathbf{p})\langle \mathbf{f}, \mathbf{Wf}\rangle\right) \qquad (4)$$

Here, $\mathbf{f}(\mathbf{p})$ is the vector $(f_1(p_1),...,f_n(p_n))$; $D(\mathbf{f}) = diag(f_1, f_2..., f_n)$; $\mathbf{W} = (w_{ij})$ is the matrix of interactions; and $\theta(\mathbf{p}) = \langle \mathbf{e}, \mathbf{f}(\mathbf{p})\rangle$, where $\langle \cdot, \cdot \rangle$ denotes inner product. Obviously, since $\langle \dot{\mathbf{p}}(t), \mathbf{e}\rangle \equiv 0$ and $f_i(0) = 0$ it follows that the simplex $\sigma_p^n$ and each of its faces are invariant sets for system (4). Note that system (4), as well as the generalized Lotka-Volterra equations [15, 16], determine the dynamics of objects with nonlinear pairwise interactions; the matrix $\mathbf{W}$ determines the structure of interactions, and the response functions determine their types. The elements of the matrix $\mathbf{W}$ are generalized interaction strengths, i.e., by analogy with the thermodynamics of irreversible processes, are "reasons" causing changes in the speed of flows [49].

*2.2 Escort distributions*

Let us rewrite Eqs. (4) as

$$\dot{\mathbf{p}} = \theta D(\mathbf{f})\left(\mathbf{Wf}\theta^{-1} - \mathbf{e}E(\mathbf{p})\right), \qquad (5)$$

where $E(\mathbf{p}) = \theta^{-2}(\mathbf{p})\langle \mathbf{f}, \mathbf{Wf}\rangle$. Using the terminology of neural networks theory [27], we refer to $E(\mathbf{p})$ as the energy function of the macrosystem under consideration. It is important to point out that in the population genetic theory this function named as population fitness [19], and in the evolutionary game theory as utility function [22]. System (5) and the structure of energy function $E(p)$ naturally determine the introduction of new additional variables:

$$x_i(p) = f_i(p_i)\theta^{-1}(p) \qquad i=1,\ldots,n. \tag{6}$$

Obviously, $x = (x_1, x_2, \ldots, x_n) \in \sigma_x^n = \{x \in \mathbb{R}^n : x_i \geq 0, e^T x = 1\}$ for $p \in \sigma_p^n$. The indices $x$ and $p$ are used in the notation of simplices in order to avoid confusion. Consider change (6) in more detail. If this is a diffeo-morphism, then it can be regarded not only as a simplifying change of variables customary in the theory of differential equations but also as the definition of a set of quantities with particular physical meaning.

The somewhat different change of variables $x = Sp(\|Sp\|_r)^{-1}$, $p \in \sigma_p^n$, where $S \geq 0$ is a nonnegative nonsingular matrix and $\|\cdot\|_r$ is the Holder $r$-norm, was considered in [15] for the classical Fischer equations (i.e., in the case $f_i(p_i) = p_i$). Under such a change, the motion of the transformed system occurs on the segment of the hypersphere $\|x\|_r = 1$ whose boundaries are determined by the relation $p \in \sigma_p^n$. In particular, this transformation has made it possible to obtain an expression for the weighted Kullback-Leibler relative entropy [15,16].

However, in the case of nonlinear pairwise interactions, the approach based on "mixing" variables leads to no substantial progress. Let us return to (6). Apparently, for the case of power functions $f_i$ of the same degree, this change was first used in 1940 in research in quantum mechanics [50]. Subsequently, in analysis of the thermodynamics of chaotic systems [51], it was proposed to use the name "escort distributions" for such distributions, which quite adequately reflects the fact of the matter. A fairly comprehensive survey of escort distributions is contained in [52]. In [48], it was shown that the problem of the existence of a one-to-one transformation into (6) in the complete space $\mathbb{R}^n$ reduces to a generalized functional Cauchy equation and has a unique solution if and only if all functions $f_i(p_i)$ are power functions of the same degree. However, considering mapping (6) on its natural domain, we can prove that this is a $C^1$ diffeomorphism on this domain.

**Theorem 1.** [53] *Under the conditions (remark 3) used for the response functions $f_i$, for $p \in \sigma_p^n$, there exists a one-to-one inverse mapping to* (6), *which is defined by*

$$p_i = f_i^{-1}(x_i) \Big/ \sum_{j=1}^n f_j^{-1}(x_j), \qquad i=1,2\ldots,n, \tag{7}$$

*where $f_i^{-1}(\cdot)$ denotes the function inverse to $f_i(\cdot)$.*

**Remark 4.** *Recall that the notation $f_i^{-1}(\cdot)$ is used for both functions inverse in the sense of function theory and functions inverse in the algebraic sense. It is always clear from the context what is meant.*

**Proof.** Suppose given a distribution $p \in \sigma_p^n$ and the related mapping $F: \sigma_p^n \to \sigma_x^n$ defined by (6). We must prove that there exists a one-to-one inverse $F^{-1}: \sigma_x^n \to \sigma_p^n$ and it is defined by (7). Let us denote the unit $n$-cube by $\mathbb{K}_y^n = \{y: 0 \leq y_i \leq 1, i = 1, \ldots, n\}$. By virtue of the assumptions made above, we have $y = f(p) \in \mathbb{K}_y^n$. Let us decompose the mapping $F$ into two stages as $F: \sigma_p^n \xrightarrow{f} \mathbb{K}_y^n \xrightarrow{g} \sigma_x^n$, where

$$g_i(y) = y_i \Big/ \sum_{j=1}^n y_j = x_i \quad i = 1, \ldots, n \tag{8}$$

The points $y$ determine a pencil of rays centered at $0 = (0, \ldots, 0)$ in $\mathbb{R}_+^n$, and relations (8) specify, in an obvious way, a system of homogeneous coordinates. Indeed, (8) implies $\dfrac{y_i}{x_i} = \dfrac{y_j}{x_i}$ for $i, j = 1, 2, \ldots, n$, i.e., $y_i = \alpha x_i$, where $\alpha \in [0, \infty)$, which determines a perspective correspondence between the rays in the pencil and the points of the simplex $\sigma_x^n$. Identifying points $y \in \mathbb{K}_y^n$ with the corresponding rays, we obtain a one-to-one correspondence between the points of the simplex $\sigma_x^n$ and segments of rays being elements of the cube $\mathbb{K}_y^n$, which leads to the diagram

$$\begin{array}{ccc} \sigma_p^n & \xrightarrow{f} & \mathbb{K}_y^n \\ g^{-1} \uparrow & & \downarrow g \\ \mathbb{K}_z^n & \xleftarrow{f^{-1}} & \sigma_x^n \end{array}$$

where $z = (z_1, \ldots, z_n)$, $z_i = f_i^{-1}(x_i)$. Since the functions $f_i$ monotonically increase on the interval [0, 1], it follows that there exist single-valued inverse functions $f_i^{-1}$ defined and increasing on the same interval. The mapping $g: \mathbb{K}_y^n \to \sigma_x^n$ determining the perspective correspondence has a one-to-one inverse $g^{-1}: \mathbb{K}_z^n \to \sigma_p^n$ as well, which proves the commutativity of diagram and completes the proof of the theorem.

*2.3. Equilibriums*

To go further, we need the following assertion.

**Statement 1.** *System* (4) *is invariant with respect to the replacement of the interaction matrix* $\mathrm{W}$ *by a perturbed matrix* $\mathrm{W}_\zeta = \left(\mathrm{W} + \mathrm{e}\zeta^T(\mathrm{p})\right)$, *where the components of the vector function* $\zeta(\mathrm{p}) = (\zeta_1(\mathrm{p}),...,\zeta_n(\mathrm{p})) : \sigma_p^n \to \mathbb{R}^n$ *are bounded on* $\sigma_p^n$.

This statement was proved in [15] for the case $\zeta(\mathrm{p}) = const$. This proof carries over in an obvious way to the case of any bounded functions $\zeta_i(\mathrm{p})$ $i = 1, 2, ..., n$. The meaning of this statement is in part clarified by the following lemma proved in [15].

**Lemma.** *Let* $\mathrm{A}$ *be a linear operator on the complex space* $\mathbb{C}^n$ *having a characteristic number* $\lambda_k(\mathrm{A})$ *with simple elementary divisor and eigenvector* $\mathrm{c}$, *and let* $\zeta \in \mathbb{C}^n$ *be an n-vector. Then the characteristic numbers of the perturbed operator* $\mathrm{B} = \mathrm{A} + \langle \bullet, \zeta \rangle \mathrm{c}$ *are preserved up to the translation* $\lambda_k(\mathrm{B}) = \lambda_k(\mathrm{A}) + \langle \mathrm{c}, \zeta \rangle$ *for any* $\zeta \in \mathbb{C}^n$.

The assertion stated above is useful because the phase space of system (4) is the standard simplex $\sigma_p^n$ contained in the subspace of $\mathbb{R}^n$ orthogonal to the unit vector $\mathrm{e}$.

In due time, this fact served as a foundation for the creation of information geometry, which is the science studying various Riemannian metrics on manifolds of probability distributions. Apparently, the first work in this direction was paper [54], in which a Riemannian metric on $\mathrm{Int}\,\sigma_p^n$ based on the Fisher information metric was introduced. At present, this metric is known as the Fisher-Rao Riemannian metric on $\mathrm{T}_p \sigma_p^n$ is given by

$$M_p(\mathrm{u}, \mathrm{v}) = \sum_{i=1}^n \frac{u_i v_i}{p_i}, \quad \mathrm{u}, \mathrm{v} \in \mathrm{T}_p \sigma_p^n,$$

where $\mathrm{T}_p \sigma_p^n$ is the tangent space in point p to $\mathrm{Int}\,\sigma_p^n$, i.e. $\mathrm{u}^T \mathrm{e} = \mathrm{v}^T \mathrm{e} = 0$. In fundamental monograph by Chentsov [55], it was proved, in particular, that this metric is a unique Riemannian metric invariant with respect to the Markov morphism. This metric usually used for classical Fisher equations. If follows in the standard way that if we deal with system (4) then a Riemannian metric is given by

$$M_f(\mathrm{u}, \mathrm{v}) = \sum_i^n \frac{u_i v_i}{f_i(p_i)}, \quad \mathrm{u}, \mathrm{v} \in \mathrm{T}_p \sigma_p^n,$$

Results of further studies on the methods of information geometry of divergence functions are given in [56]. Statement 1 proved above makes it possible to investigate system (4) of probability distributions evolution equations with taking into account the fact that we deal with the restriction of system (4) to the simplex $\sigma_p^n$.

We proceed to consider questions related to the existence of a nontrivial equilibrium point $\hat{p} \in \text{Int}\,\sigma_p^n$ for system (4). In [15, 48], it was proved that $\hat{p} > 0$ exists if and only if all components of the vector $W^{-1}e$ are of the same sign. It is clear already from this result that properties of the inverse interaction matrix play a substantial role in the evolution of macrosystems. Statement 1 implies that the coordinates of a nontrivial equilibrium point remain invariable $\forall \zeta \in \mathbb{R}^n$. Let us prove this directly.

**Statement 2.** [53] *If system* (4) *has a nontrivial equilibrium point* $\hat{p} \in \text{Int}\,\sigma_p^n$, *then*

$$\frac{\hat{f}}{\langle \hat{f}, e \rangle} = \frac{W^{-1}e}{\langle e, W^{-1}e \rangle} = \frac{W_\zeta^{-1}e}{\langle e, W_\zeta^{-1}e \rangle} = \hat{x} \qquad \forall \zeta(p) \in \mathbb{R}^n \qquad (9)$$

**Proof.** We use the following notation:

$W^{-1}e = a$, $\langle e, W^{-1}e \rangle = b$, i.e. $\hat{x} = \dfrac{a}{b}$. Using an expression for the matrix $W_\zeta^{-1}$ given in [57] (p. 139), we obtain

$$\left(W + e\zeta^T\right)^{-1} = W^{-1} - \frac{a\zeta^T}{1 + \zeta^T a} W^{-1}$$

therefore, $W_\zeta^{-1}e = \dfrac{a}{1 + \zeta^T a}$. Multiplying this relation by $e^T$ on the left, we obtain

$$e^T W_\zeta^{-1} e = \frac{e^T a}{1 + \zeta^T a} = \frac{b}{1 + \zeta^T a},$$

i.e., $\dfrac{W_\zeta^{-1}e}{e^T W_\zeta^{-1}e} = a/b = \hat{x}$, $\forall \zeta(p) \in \mathbb{R}^n$, which completes the proof of the statement.

**Corollary 1.** *The escort distribution equilibrium value $\hat{x}$ depends only on the properties of the inverse interaction matrix, and the equilibrium vector $\hat{p}$ is determined from the systems* (7) *and* (9):

$$\hat{p}_i = f_i^{-1}(\hat{x}_i) \Big/ \sum_{j=1}^n f_j^{-1}(\hat{x}_j), \qquad i = 1, \ldots, n \qquad (10)$$

The distribution (10) is sometimes called the Gibbs distribution. The normalization constant is determined as usual by

$$Z = \sum_{j=1}^n f_j^{-1}(\hat{x}_j) \qquad (11)$$

and is conventionally called the partition function.

*2.4. Energy-like Lyapunov-Meyer functions*

Following the stability postulate, we state the following theorem, obtained in [53]

**Theorem 2.** *If there exists a vector $\zeta$, for which $W_\zeta = W_\zeta^T$, then the energy function of the system*

$$E_\zeta(p) = \langle f(p), W_\zeta f(p) \rangle \theta^{-2}(p) \tag{12}$$

*is its Lyapunov-Meyer function.*

The proof of this theorem is based on the fact that, under the assumptions of the theorem, system (5) can be written in the gradient form

$$\dot{p} = D(f(p)) \nabla E_\zeta(p),$$

where $\nabla E_\zeta(p) = (\partial E_\zeta(p)/\partial p_1, ..., \partial E_\zeta(p)/\partial p_n)$ is the gradient operator and $D(f(p))$ is a diagonal matrix with positive elements. We point out the obvious fact that the energy function contains no information about the equilibrium of the system (5), and the landscape determined by this function makes it possible to perform a complete qualitative study of the system trajectories behavior in its phase space, $\sigma_p^n$, including its faces. We emphasize that this result is a substantial generalization of the fundamental theorem of natural selection [19].

**Remark 5.** *The symmetry conditions on the interaction matrix in Theorem 2 are related in an obvious way to the well known Onsager reciprocity relations* [49].

Theorem 2 has the following useful corollary.

**Corollary 2.** *If the interaction matrix $W$ is tridiagonal, then there exists a vector $\zeta \in \mathbb{R}^n$ such that the energy function $E_\zeta$ of the system is a Lyapunov-Meyer function.*

**Proof.** The system of equations $W + e\zeta^T = W^T + \zeta e^T$ in the case of a tridiagonal matrix $W$ gives $(n-1)$ linear equations determining the components of the vector $\zeta$, which leaves some freedom in the choice of the required symmetrizing vector $\zeta$.

Thus, if pairwise interaction occurs only between "nearest neighbors," then the evolution of the macrosystem is of gradient type.

Also from (9) and (12) we can receive the expression for equilibrium energy

$$\hat{E} = E(\hat{p}) = \langle e, W^{-1} e \rangle^{-1} \tag{13}$$

Note, that $\hat{E}$ depends only from interaction matrix.

*2.5. Entropy-like Lyapunov-Meyer functions*

**Theorem 3.** [53] *If system (4) has a nontrivial equilibrium point $\hat{p} \in \text{Int}\, \sigma_p^n$ and the matrix $(W_\zeta^T + W_\zeta)$, $\forall \zeta \in \mathbb{R}^n$ has $n-1$ negative characteristic numbers, then the function*

$$H(\mathrm{p}) = \sum_{i=1}^{n} \int_{\hat{p}_i}^{p_i} \frac{\hat{f}_i dx}{f_i(x)} + C, \quad (14)$$

where $C$ is a constant, is a Lyapunov-Meyer function for system (4) on $\mathrm{Int}\,\sigma_p^n$, and the energy function $E(\mathrm{p})$ of the system attains its maximum value $E(\hat{\mathrm{p}})$ as $t \to \infty$.

**Proof.** For the derivative of the function $H(\mathrm{p})$ along the trajectories of system (4), we have

$$\dot{H}(\mathrm{p}) = \hat{\mathrm{f}}^T \left( \mathrm{Wf} - \mathrm{e}\theta^{-1} \langle \mathrm{f}, \mathrm{Wf} \rangle \right) \quad (15)$$

Let us rewrite (15) in terms of the escort distribution $x(t)$. Taking into account the relation $x^T \mathrm{e} = 1$, we obtain

$$\dot{H}(\mathrm{p}) = \hat{\theta}\theta \left( \hat{x}^T \mathrm{W} x - x^T \mathrm{W} x \right) \quad (16)$$

Thus, we must prove that the hyperplane $\tilde{E}(x) = \hat{x}^T \mathrm{W} x$ is tangent to the surface $E(x) = x^T \mathrm{W} x$ at the point $x = \hat{x}$. Let us introduce the auxiliary variables $v_i = x_i - \hat{x}_i$, $i = 1,...,n$. The vector $v$ belongs to the "displaced" simplex $\sigma_v^n = \{v : \mathrm{e}^T v = 0,\ -\hat{x}_i < v_i < 1 - \hat{x}_i\}$, which is the tangent space to $\mathrm{Int}\,\sigma_x^n$. After simple transformations, we obtain

$$\dot{H}(\mathrm{p}) = -\hat{\theta}\theta \langle v, \mathrm{W}v \rangle. \quad (17)$$

According to the conditions of the theorem and Kingman's result [58], the quadratic form $\langle v, \mathrm{W}v \rangle$ is negative definite on $\sigma_v^n$, i.e., $\dot{H}(\mathrm{p}) > 0$ for $\mathrm{p} \neq \hat{\mathrm{p}}$, which proves the first part of the theorem. Relation (15) with (5) taken into account can be rewritten as

$$\dot{H}(\mathrm{p}) = \theta\hat{\theta} \left( \tilde{E}(\hat{\mathrm{p}}) - E(\mathrm{p}) \right) \geqslant 0, \quad (18)$$

where $\tilde{E}(\hat{\mathrm{p}}) = \langle \hat{\mathrm{f}}, \mathrm{wf} \rangle \hat{\theta}^{-1}\theta^{-1}$, which completes the proof of the theorem.

**Remark 6.** *As is easy to see, relation (18) determines the entropy production rate in the system. In [46], il was shown that if the interaction matrix W satisfies the condition $\mathrm{W}^T\mathrm{W}^{-1}\mathrm{e} = \mathrm{e}$, then $\tilde{E}(\hat{\mathrm{p}}) = E(\hat{\mathrm{p}})$, which sometimes simplifies the consideration of relation (18).*

**Remark 7.** *It is obvious that entropy function (14) has the decomposable form in the sense of [2].*

Consider properties of the function $H(\mathrm{p})$. We begin with a quotation from [55]: "The most important information invariant of a pair of probability measures is relative entropy. ... This characteristic is asymmetric; therefore, we call it the information deviation of one distribution from the other... ." Chentsov's book [55] was written in 1972. Since then, many terminological changes have occurred in this area; they are summarized in the recently published *Dictionary of Distances* [59]. We shall use the term *relative entropy,* or the function of information divergence.

The function $H(\mathrm{p})$ with $\mathrm{p} \in \sigma_p^n$ attains its maximum value at the point $\mathrm{p} = \hat{\mathrm{p}}$, where $H(\hat{\mathrm{p}}) = C$; i.e., $H(\mathrm{p}) < C$ $\forall \mathrm{p} \neq \hat{\mathrm{p}}$, the covector $H(\mathrm{p})$ is $\nabla H(\mathrm{p}) = \left(\hat{f}_i / f_i(p_i)\right)$, and the Hessian is $\mathcal{H}(H(\mathrm{p})) = -diag\left(\hat{f}_i \dfrac{\partial f_i}{\partial p_i} f_i^{-2}(p_i)\right)$. Since $\partial f_i / \partial p_i > 0$, it follows that the Hessian is negative definite, and the function $H(\mathrm{p})$ is concave on $\sigma_p^n$. Thus, $-H(\mathrm{p})$ can be regarded as relative entropy, or the function of information divergence between the distributions p and $\hat{\mathrm{p}}$.

Note that the relative entropy $H(\mathrm{p})$ introduced by us depends on the interaction matrix only via the equilibrium point $\hat{\mathrm{p}}$ of the system and is largely determined by the form of the response functions. This allows us to use relation (14) for solving the "inverse" problem of finding the response functions corresponding to a given entropy function $H(\mathrm{p})$.

In the **results** section we use Theorem 3 to construct entropy functions and "distances" between probability distributions. First, we note that, if $G: \mathbb{R}^1 \to \mathbb{R}^1$ is a monotonically increasing smooth function and $C_1$, $C_2$ and $C_3$ are the numbers, then the function

$$H_G = C_1 G(C_2 H(\mathrm{p})) + C_3 \qquad (19)$$

is also a Lyapunov-Meyer function for system (4). Clearly, the domain of $G$ coincides with the range of $(H(\mathrm{p}) + C)$, $\mathrm{p} \in \mathrm{Int}\sigma_p^n$.

*2.6. Distanse measures*

Following Shannon information measure idea let us introduce $S(\mathrm{p})$ as negative entropy defined by:

$$S(\mathrm{p}) = -H(\mathrm{p}) = -\sum_{i=1}^{n} \int_{\hat{p}_i}^{p_i} \dfrac{\hat{f}_i dx}{f_i(x)}, \quad \mathrm{p} \in \mathrm{Int}\sigma_p^n \qquad (20)$$

It is obvious by the previous statement that the function $S(\mathrm{p})$ is defined on the $\mathrm{Int}\sigma_p^n$, and we have to use directional derivative of $S$ for its analysis. However in this case it is easy to use the following evident identity

$$\lambda \sum_{i=1}^{n} \int_{\hat{p}_i}^{p_i} dx = \lambda \sum_{i=1}^{n} (p_i - \hat{p}_i) \equiv 0 \quad \forall p, \hat{p} \in \mathrm{Int}\sigma_p^n$$

and using (20) we clearly have, the following relationship

$$S(\mathrm{p}) = -\sum_{i=1}^{n} \int_{\hat{p}_i}^{p_i} \left( \frac{\hat{f}_i dx}{f_i(x)} + \lambda \right) dx \quad \forall \lambda \in \mathbb{R}^1$$

Since, by the evident equality $\partial S(\mathrm{p})/\partial p_i = 0$, on $p = \hat{p}$, we have that $\lambda = -1$. It is clear, that $\lambda$ is a type of Lagrange multiplier. Therefore:

$$S(\mathrm{p}) = \sum_{i=1}^{n} \int_{\hat{p}_i}^{p_i} \left( 1 - \frac{\hat{f}_i}{f_i(x)} \right) dx \quad \mathrm{p} \in \mathrm{Int}\sigma_p^n \tag{21}$$

The gradient of $S(\mathrm{p})$ on $\mathrm{p} \in \mathrm{Int}\sigma_p^n$ is given by vector of partial derivatives

$$\nabla S(\mathrm{p}) = \left( 1 - \frac{\hat{f}_i}{f_i(p_i)} \right) \tag{22}$$

with strictly monotonically increasing each component. Hessian of $S(\mathrm{p})$ on $\mathrm{Int}\sigma_p^n$ is $\mathcal{H}(\mathrm{p}) = diag\left( \hat{f}_i \frac{\partial f_i}{\partial p_i} f_i^{-2}(p_i) \right)$. Since $\partial f_i/\partial p_i > 0$ it follows that the Hessian is positive definite, and the function $S(\mathrm{p})$ according to definition is convex on $\mathrm{Int}\sigma_p^n$. As observed by Ball and Chen [60] entropy and convexity have played an important role in many areas of mathematics. Continuing this line of reasoning, we see that $S(\mathrm{p})$ is an distance measure between probability distributions $\mathrm{p}$ and $\hat{\mathrm{p}}$.

Another very well known definition of convexity Jensen [61], Hardy et al. [62] is the next Jensen inequality

$$S(\mathrm{p}) - S(\mathrm{q}) - \langle \nabla_q S(\mathrm{q}), \mathrm{p} - \mathrm{q} \rangle \geq 0 \quad \mathrm{p}, \mathrm{q} \in \mathrm{Int}\sigma_p^n \tag{23}$$

It is obvious that the expression from left-side inequality is so called Bregman divergences usually denote by $B_S(\mathrm{p}, \mathrm{q})$. This name was first given the by Censor and Lent [63]. Bregman divergence or Bregman distance [64] is similar to a metric, but does not satisfy the triangle inequality nor symmetry. Using inequality (23) we can receive new weighted distance measure between two probability distributions $\mathrm{p}$ and $\mathrm{q}$.

$$B_S(\mathrm{p}, \mathrm{q}) = \sum_{i=1}^{n} \left( \int_{p_i}^{q_i} \frac{\hat{f}_i dx}{f_i(x)} + \frac{\hat{f}_i}{f_i(q_i)} (p_i - q_i) \right) \geq 0$$

Evidently that $B_S(\mathrm{p}, \mathrm{q}) = 0$ for $\mathrm{p} = \mathrm{q}$, and $B_S(\mathrm{p}, \hat{\mathrm{p}}) = S(\mathrm{p})$.

Another measures of information was proposed by Renyi in 1960 [65]. Today this measures is known as f-divergences. In 1963 Csiszar [66] and Morimoto [67] discovered this functions independently and now it is more known as Csizar divergences

$$D_f(p,q) = \sum_{i=1}^{n} q_i h\left(\frac{p_i}{q_i}\right)$$

where $h$ is a convex function on $(0,\infty)$ such that $h(1)=0$

Note that this function have found numerous application during the past years, and many works are devoted to an analysis of $D_f(p,q)$ [68]. An analysis of relations between the functions (14) and (21) introduced by us and f-divergences will be the object of next paper.

*2.7. Legendre-Donkin-Fenchel transformations*

The Legendre-Donkin-Fenchel (LDF) transformation is a mathematical concept of great significance to thermodynamics, mechanics and probability theory is named after Legendre [69] Donkin [70] and Fenchel [71]. LDF-transformation is a way for representing a function in terms of its first derivative.

Let us denote $l_i = \partial S(p)/\partial p_i$. Since $l_i$ is strictly monotonically increasing for $p_i$, this makes it clear that $l_i$ and $p_i$ are mutually reciprocal. The transformation from vector $p$ to vector $l$ is called LDF-transformation. We can find a convex function $L$ of $y$ defined by

$$L(l) = \sum p_i l_i - S(p) \qquad (24)$$

when $l$ and $p$ are respective coordinates of the same point, and the inverse transformation from $l$ to $p$ is given by the gradient

$$p = \nabla L(l) \qquad (25)$$

Concerning (22) we clearly have

$$l_i = \frac{\partial S(p)}{\partial p_i} = 1 - \frac{\hat{f}_i}{f_i(p_i)}$$

and

$$p_i = f_i^{-1}\left(\frac{\hat{f}_i}{1-l_i}\right) = \frac{\partial L}{\partial l_i}$$

for $l_i \in (-\infty, 1-\hat{f}_i)$. We also denote $\partial L/\partial l_i = \varphi(l_i)$ and after simple counting we have the following relationship

$$L(l) = \sum_{i=1}^{n}\left(l_i \varphi(l_i) + \int_{0}^{\varphi(l_i)}\left(\frac{\hat{f}_i}{f_i}-1\right)dx\right) \qquad (26)$$

After simple computation one can check that $\partial L/\partial l_i = p_i$. Another very well known form of LDF-transformation, together with (24) is

$$L(l) = \max_{p}(\langle p, l \rangle - S(p)) \tag{27}$$

Follow Amari [68,72] and definition (24) one can define a divergence function (distance measure) between two points $p$ and $q$ on $\sigma_p^n$ in the dual coordinates. This divergence function will be similar to the squared Riemannian distance. Additional new recent information on LDF-transformation is presented in Nielsen [73] and Zia el al [74] papers.

## 3. Examples

*3.1. Boltzmann entropy*

Consider the Boltzmann-Shannon entropy $H(\mathrm{p}) = -\sum_{i=1}^{n} p_i \ln p_i$. Taking into account the background of the system (4) it is clear that the interaction matrix $W$ is stochastic, i.e. $W\mathbf{e} = \mathbf{e}$ and the uniform distribution $\hat{\mathrm{p}} = \mathbf{e} n^{-1}$ must be a nontrivial equilibrium point of the system, and all response functions $f_i(\cdot)$ are the same. From the equation

$$-\sum_{i=1}^{n} p_i \ln p_i = \sum_{i=1}^{n} \frac{1}{n} \int_{n^{-1}}^{p_i} \frac{dx}{f_i(x)} + C$$

using the notation of Theorem 1 for nonlinear response functions, we obtain

$$y_i = f_i(p_i) = (1 - \ln p_i)^{-1}, \quad i = 1, \ldots, n. \tag{28}$$

Obviously that on the interval [0, 1], functions (28) satisfy the conditions specified in Remark 3 and this is convex functions. It can be shown in the usual way that the inverse functions

$$f_i^{-1}(y_i) = \exp(1 - 1/y_i), \quad y \in \mathbb{K}_y^n, \quad i = 1, \ldots, n \tag{29}$$

It is clear, that (29) is concave functions. If we consider equations of evolution (4) with an arbitrary interactions matrix $W$ satisfying the condition of Theorem 3, then instead of Jaynes [75] maximum-entropy principle we can receive the equilibrium probability distribution from (9) and (10):

$$\hat{\mathrm{p}}_i = \exp\left(-\frac{1}{\hat{x}_i}\right) \Big/ Z$$

Recall that $\hat{x}_i = \hat{E} W^{-1} \mathbf{e}$, so $\hat{x}$ is independent from the type of interactions.

Of most interest in this example are, certainly, functions (28) and (29), which make it possible to not only apply the developed theory of pairwise interaction system but also consider the problem of finding a possible equivalence relation between the Boltzmann kinetic equations based on pairwise interactions and system (4).

*3.2. Power-law entpopies*

Suppose that the matrix W satisfies the conditions of Theorem 3. As the functions $f_i(p_i)$, we consider power-law function

$$f_i(p_i) = p_i^{1-\alpha}, \quad \alpha < 1, \quad i = 1,...,n \tag{30}$$

Performing calculation we obtain from (14)

$$H(\mathrm{p}) = \sum_{i=1}^{n}\left(\frac{p_i^\alpha \hat{p}_i^{(1-\alpha)}}{\alpha} - \frac{\hat{p}_i}{\alpha}\right) + C \tag{31}$$

If $\hat{\mathrm{p}} = \mathbf{e}n^{-1}$, then setting $C = \dfrac{1}{\alpha}$, we received power-law entropy function

$$H(\mathrm{p}) = \frac{n^{(\alpha-1)}}{\alpha} \sum_{i=1}^{n} p_i^\alpha \tag{32}$$

Next we apply (19) and show how we can receive, for example, Renyi and Tsallis entropy functions.

**Renyi entropy function.** Setting in (19) $C_2 = \alpha/n^{(\alpha-1)}$, $C_1 = (1-\alpha)^{-1}$, $G(\cdot) = \ln(\cdot)$ we obtain the Renyi entropy function [65]:

$$H_R = (1-\alpha)^{-1} \ln\left(\sum_{i=1}^{n} p_i^\alpha\right) \tag{33}$$

**Tsallis entropy function.**

The reader will have no difficulty in showing the same way that Tsallis entropy function [76]:

$$H_T = \left(\sum_{i=1}^{n} p_i^\alpha - 1\right)\bigg/(1-\alpha) \tag{34}$$

Recall that function (34) was introduced by Tsallis for the purpose of constructing the thermodynamics of nonextensive systems, and the parameter $\alpha$ is a nonextensivity measure of the system. The creation of such a thermodynamics is necessary because large systems with so called long-range interactions cannot be described by the Boltzmann thermodynamics.

Note that expression (32) is valid as LMF for (4) only if $0 < \alpha < 1$. The lower bound is determined by the requirement that the improper integrals in (14) must converge and the upper bound follows from the equalities $f_i(0) = 0$.

*3.3. Kullback-Leibler distance measure*

In the following example, we consider one of the best known measures of distance between probability distributions. Suppose that the matrix W satisfies the conditions of Theorem 3. We set $f_i(p_i) = p_i$ for $i = 1, 2, ..., n$. Then (14) gives the Kullback-Leibler relative entropy [77]

$$H_{\hat{p}}(p) = \sum_{i=1}^{n} \hat{p}_i \ln(p_i / \hat{p}_i)$$

First time, this function was used as Lyapunov function to analyze the stability of equilibrium points of replicator systems in [78]. Note that, since the functions $f_i$ are linear in this case, we deal with the classical system of replicator equations. A detailed analysis of such equations was made in [15]; in particular, it was shown that, for such systems, we can remove the constraint $W = W^T$ and obtain results similar to those obtained above for matrices of the type $W = D_1 B D_2$, where $D_1$ and $D_2$ are diagonal matrices, $D_1 D_2 > 0$, and $B = B^T$, i.e., for the class of so-called diagonally symmetrizable matrices. A necessary and sufficient condition for a real matrix $W$ to be diagonally symmetrizable was obtained in [16]. It is so far unclear whether the constraints $W = W^T$ can be removed in the case of nonlinear functions $f_i$.

## 4. Discussion

It is seen from the examples given above that many (and practically all) known entropy functions and distance measures my be obtain from entropy-like Lyapunov functions (14) and (19). We also emphasize that there exists a relation between the derivative of the function $H(p)$, which can be interpreted as a generalized entropy, and the function $E(p)$, which is often considered as an analog of the neural networks energy or population fitness. This relationship for entropy production was establihed by Pykh [44,53] for corresponding types of interactions matrix and has the next form:

$$\dot{H}(p) = \theta \hat{\theta} \left( \tilde{E}(\hat{p}) - E(p) \right) \geqslant 0$$

Note that it was Ilya Prigogine who the first pointed out [5] the importance of the relationship between Lyapunov functions and entropy.

The approach developed in this paper makes it possible to associate each entropy functions with some set of dynamical systems which explicitly depend on the character of the interactions between the objects under examination, which, in the author's opinion, may be essential for solving particular problems.

Note that obtained results can be used in two ways.
1. The response functions can be found for previously known entropy characteristics. Relevant examples for the Shannon and Tsallis entropy were given in [79]. It is easy to show that this approach yields response functions for all the entropy characteristics proposed in [80].

2. New entropy characteristic can be derived from some functions obeying the conditions stated for response functions. As an example, we consider the logistic function, which is widely applied in mathematical ecology and economics [46].

In conclusion, we emphasize that system (4) is based on the acceptance of only two hypotheses about the structure and the type of nonlinear pairwise interactions. However, results of the mathematical study of this system make it possible to obtain not only the classical Boltzmann-Shannon entropy but also any other known parametric entropy function.

We also mention that all results stated above were obtained by formally analyzing systems of generalized replicator equations, which arise in very diverse fields of natural sciences and, therefore, can serve as a basis for finding analogies between these domains of natural sciences.

Let us point out also that construction of LMF for general nonlinear systems is of great theoretical and practical interests [81]. If this function exists, that means that the system under considerations is a gradient system [82].

The reader may consider this paper as a step in the development of nonlinear pairwise interactions dynamical systems theory.

**Conflits of Interest**

The author declare no conflict of interest.